# Mass enhancement and metal-nonmetal transition driven by *d-f* hybridization in perovskites La$_{1-x}$Pr$_x$CuO$_3$

H. Takahashi[1,2], M. Ito[3], J. Fujioka[4], M. Ochi[5,6], S. Sakai[7], R. Arita[3,7], H. Sagayama[8], Y. Yamasaki[9], and S. Ishiwata[1,2]

[1]*Division of Materials Physics and Center for Spintronics Research Network (CSRN), Graduate School of Engineering Science, Osaka University, Osaka 560-8531, Japan*
[2]*Spintronics Research Network Division, Institute for Open and Transdisciplinary Research Initiatives, Osaka University, Yamadaoka 2-1, Suita, Osaka, 565-0871, Japan*
[3]*Department of Applied Physics, the University of Tokyo, Bunkyo-ku, Tokyo 113-8656, Japan*
[4]*Department of Materials Science, University of Tsukuba, 1-1-1 Tennodai, Tsukuba, Ibaraki 305-8573, Japan*
[5]*Department of Physics, Osaka University, Machikaneyama-cho, Toyonaka, Osaka 560-0043, Japan*
[6]*Forefront Research Center, Osaka University, Machikaneyama-cho, Toyonaka, Osaka 560-0043, Japan*
[7]*RIKEN Center for Emergent Matter Science, 2-1 Hirosawa, Wako, 351-0198, Japan*
[8]*Institute of Materials Structure Science (IMSS), High Energy Accelerator Research Organization (KEK), Tsukuba, Ibaraki 305-0801, Japan*
[9]*National Institute for Materials Science (NIMS), Tsukuba, Ibaraki 305-0047, Japan*

We report the large electron-mass enhancement and the metal to nonmetal transition upon the Pr doping in perovskite-type La$_{1-x}$Pr$_x$CuO$_3$. With increasing the Pr content $x$ around 0.6, the LaCuO$_3$-type three-dimensional structure with trivalent Cu ions changes to the quasi-one-dimensional structure with nearly divalent Cu ions, which accompanies significant changes in the electronic properties. Based on the resistivity, optical conductivity, specific heat measurements and the first-principles calculations, we discuss the formation of a nearly localized nonmetallic state stabilized by the hybridization between Cu $3d$, O $2p$, and Pr $4f$ orbitals in the quasi-one-dimensional lattice. The present perovskite-type cuprates offer a unique opportunity to explore novel quantum phases of correlated electrons in low-dimensional lattice, where the spin/charge/orbital degrees of freedom of $A$- and $B$-site ions are entangled.

Perovskite-type transition-metal oxides $ABO_3$ have been extensively explored as functional materials showing a rich variety of magnetic and electronic properties [1]. The rich electronic functions derive typically from the strongly correlated $d$ electrons in the three-dimensional $B$-O lattice. This is exemplified by $A$MnO$_3$ with a giant magnetoelectric effect [2,3] and $A$NiO$_3$ with a metal-insulator transition [4,5], in which the $A$ site is occupied by the trivalent rare-earth ions. On the other hand, it has been demonstrated that the charge and orbital degrees of freedom of $A$-site ions can be involved in the electronic properties of perovskite oxides, when the $B$ site is occupied by late $3d$ transition-metal ions with high valence state. For instance, (La,Bi)NiO$_3$ and (La,Sr)Cu$_3$Fe$_4$O$_{12}$ show gigantic negative thermal expansion due to the inter-site charge transfer between Bi and Ni, and Cu and Fe, respectively [6–11].

Although considerable efforts have been paid for perovskite oxides with various $3d$-transition-metal ions, less has been known about the perovskite-type cuprates $A$CuO$_3$, which can be regarded as the three-dimensional counterparts of the high-$T_c$ cuprates with layered structure. The scarcity of perovskite-type $A$CuO$_3$ presumably reflects the fact that $A$CuO$_3$ with unusually high-valence Cu$^{3+}$ ions are typically difficult to obtain, while oxygen-deficient perovskites YBa$_2$Cu$_3$O$_{6+x}$ with superconducting CuO$_2$ planes have been extensively synthesized at ambient pressure [12]. So far, perovskite-type cuprates without oxygen deficiency have been reported for LaCuO$_3$ and its substitution of La by Nd [13–18]. While La$_{1-x}$Nd$_x$CuO$_3$ shows an indication of the mass enhancement with increasing $x$ up to 0.6, the system keeps the rhombohedral structure and the metallic behavior down to the lowest temperature [16]. It is presumable that the introduction of magnetic rare-earth ions into the $A$-site of $A$CuO$_3$ provides novel quantum phases inherent to the inter-site charge transfer and the $d$-$f$ hybridization.

Here we discover the $A$-site dependent large mass enhancement and metal-nonmetal transition in perovskite-type cuprates, La$_{1-x}$Pr$_x$CuO$_3$, which can be associated with the intersite charge transfer, $A^{3+}$Cu$^{3+}$O$_3$ → $A^{(4-\delta)+}$Cu$^{(2+\delta)+}$O$_3$. The $A$-site dependent transition accompanies the significant changes in both crystalline and electronic structures as the three-dimensional rhombohedral structure in a metallic state ($x < 0.6$) to the quasi-one-dimensional orthorhombic structure in a conductive but nonmetallic state ($x > 0.6$). The emergence of the nonmetallic state in La$_{1-x}$Pr$_x$CuO$_3$ is discussed in terms of a strong coupling between the $4f$ electrons of Pr$^{3+/4+}$ ion and the $3d$ electrons of Cu$^{2+/3+}$ ion.

Polycrystalline samples of La$_{1-x}$Pr$_x$CuO$_3$ were prepared as described in the previous report for PrCuO$_3$ [19] (for detailed experimental and theoretical methods, see Supplemental Material [20]). Figure 1(b) shows the powder x-ray diffraction (XRD) patterns of La$_{1-x}$Pr$_x$CuO$_3$ with selected compositions of $x$ = 0, 0.5, 0.6, 0.7, 0.9, and 1.0, indicating the emergence of the $A$-site dependent structural transition around $x$ = 0.6. The diffraction patterns were indexed with a rhombohedral ($R\bar{3}c$) unit cell for $x$ = 0 and 0.5 and orthorhombic ($Pbnm$) unit cell for $x$ = 0.7, 0.9, and 1.0. The diffraction pattern of $x$ = 0.6 contains the peaks of both structures, implying that this composition is located near the first-order structural phase boundary between them as shown in Fig. S1 (see Supplemental Material).

For the selected compounds ($x$ = 0, 0.5, 0.7, 1), the crystal structure was refined by the Rietveld analysis for the x-ray diffraction data as shown in Fig. S2 (see Supplemental Material). Figure 1(c) shows the variation of the unit cell volume divided by $Z$, the number of formula units per unit cell, as a function of the Pr content $x$. In the rhombohedral phase with $x$ less than 0.6, the unit cell volume decreases monotonically with increasing $x$, reflecting the decrease in the averaged ionic radius of the $A$-site ion. As $x$ exceeds 0.6, the rhombohedral structure is replaced by the highly distorted PrCuO$_3$-type structure, accompanying a jump in the unit cell volume by 7 %. Considering the fact that the unit cell volume of perovskite oxides is strongly affected by the $B$-O bond length, such a huge volume increase upon the structural change implies the sudden decrease in the valence of the Cu ion as a consequence of the intersite charge transfer. This conjecture is confirmed by the bond-valence-sum (BVS) calculation for Cu in La$_{1-x}$Pr$_x$CuO$_3$, as shown in Fig. 1(d). The estimated Cu valence in the composition with $x$ below 0.6 is close to +3, whereas that for $x$ above 0.6 is in the range of +2.2 to +2.4. The nearly divalent nature of Cu ions is consistent with the cooperative Jahn-Teller distortion of the CuO$_6$ octahedra, which lifts the degeneracy of the $e_g$ orbitals. This result has been supported also by the x-ray absorption near edge structure (XANES) measurements around the Cu K edge [19]. Whereas the intersite charge transfer with the increment of Pr content in La$_{1-x}$Pr$_x$CuO$_3$ reminds us of that in Bi$_{1-x}$La$_x$NiO$_3$ with the increment of Bi content [9], there exists a distinct difference between them reflecting the presence of spin and orbital degrees of freedom in the $A$-site of La$_{1-x}$Pr$_x$CuO$_3$.

The electronic state of La$_{1-x}$Pr$_x$CuO$_3$ changes significantly upon the $A$-site-dependent structural transition around $x = 0.6$. Figure 2(a) shows the temperature dependence of the electrical resistivity for La$_{1-x}$Pr$_x$CuO$_3$. Pr doping on LaCuO$_3$ increases the absolute value of electrical resistivity, and the resistivity at 2 K increases by an order of 3 upon the structural transition between $x = 0.6$ and 0.7 as shown in Fig. 1(e). La$_{1-x}$Pr$_x$CuO$_3$ with $x$ larger than 0.6 shows nonmetallic temperature dependence, implying that the charge carriers are in an incoherent state. The significant enhancement of resistivity associated with the structural transition to the quasi-one-dimensional structure presumably suggests the formation of incoherent bands near the Fermi energy. In addition, the facts that the crystal structure is highly anisotropic for $x > 0.6$ and the sample is polycrystalline, making the resistivity susceptible to scattering at grain boundaries. The $x$-dependent change in the electronic state of La$_{1-x}$Pr$_x$CuO$_3$ can be found in the optical conductivity. The photon energy dependence of optical-conductivity spectra is shown in Fig. 2(c). A clear Drude-like peak is observed below 0.2 eV for $x = 0$, which is almost absent for $x > 0.6$. However, in contrast to the Mott insulator of PrCuO$_{2.5}$, which has an energy gap of ~ 0.5 eV, a finite intensity is observed at low energies down to 0 eV for $x > 0.6$. This is consistent with the fact that the electrical resistivity of PrCuO$_{2.5}$ is three orders of magnitude larger than that of PrCuO$_3$ at room temperature as shown in Fig. S3 (see Supplemental Material [20]). The several hump-like structures below 2 eV can be ascribed to $d$-$d$ transitions inherent to the heavily distorted CuO$_6$ octahedra. These results indicate that the compounds with $x > 0.6$ have incoherent bands near the Fermi energy, which is consistent with the nonmetallic temperature dependence of the resistivity.

Figure 2(b) exhibits the temperature dependence of the magnetic susceptibility measured under a magnetic field of 0.1 T in a field cooling run for $x = 0$, 0.6, 0.7, and 1.0 (the magnetic susceptibility of LaCuO$_3$ was taken from Ref. [21]). For $x > 0.6$, the Curie-Weiss-like behavior is observed above about 50 K. The effective Bohr magnetons number $P_{\text{eff}}$ was evaluated by the Curie-Weiss law fitting between 100 and 300 K with the formula $M/H = xC_w/(T-\theta)+\chi_0$ as shown in Fig. S4 (see Supplementary Material [20]), where $C_w$, $\theta$, and $\chi_0$ are the Curie constant, Weiss temperature, and temperature-independent background contributions mainly from Pauli paramagnetism, respectively. The $P_{\text{eff}}$ decreases from 3.44 to 2.68 $\mu_B$/Pr with a change in $x$ from 0.6 to 1.0 (all data are listed in Table 1), suggesting that the valence of Pr tends to increase from +3 (theoretical value of $P_{\text{eff}} = 3.58$ $\mu_B$/Pr) to +4 ($P_{\text{eff}} = 2.54$ $\mu_B$/Pr). Here, we assume that the localized moment of Pr ions dominates the Curie-Weiss behavior. The $x$-dependent change in $P_{\text{eff}}$ is consistent with the change in the valence of Pr ions expected from the BVS calculations (Table S5 in Supplementary Material shows the BVS of La/Pr site, where the estimated La/Pr valence in the composition with $x$ below 0.6 is close to +3, and that for $x$ above 0.6 is close to +4). Whereas the Weiss temperature is negatively large as -35 K and -55 K, La$_{1-x}$Pr$_x$CuO$_3$ exhibits no magnetic transition down to 2 K. This result implies that the magnetic Pr sublattice becomes effectively one dimensional, given that the magnetic interactions between Pr ions are mediated by the Cu-O sublattice.

To further get insight into the electronic states, we measured the temperature dependence of the specific heat for selected compositions. The specific heat also depends on the Pr content $x$, as shown in Fig. 3(a) without offset. LaCuO$_3$ exhibits nonmagnetic metallic behavior with a small electronic specific heat coefficient $\gamma$ (~ 5.0 mJ/mol K$^2$). As $x$ increases above 0.7, $C/T$ shows a substantial upturn at low temperatures, as observed in oxides with Pr$^{4+}$ ions. This upturn of $C/T$ implies the schottky anomaly of Kramers doublet for Pr$^{4+}$ ions. To confirm this presumption, we evaluate the entropy in PrCuO$_3$ by subtracting the $C/T$ of LaCuO$_3$, which can be regarded as a background mainly from lattice contribution, from that of PrCuO$_3$, as shown in Figs. 3(b) and 3(c) (the detailed analysis is shown in Fig. S5). The entropies estimated from the background-subtracted $C/T$ are found to approach $R$ln2, supporting the presence of a high percentage of Pr$^{4+}$ ions with Kramers ground-state doublet in PrCuO$_3$. The temperature dependence of $C/T$ at $H = 0$ T shows no peak structure down to 2 K, presumably because the energy split of Kramers doublet is smaller than the lowest measurable temperature of 2 K [22]. The emergence of the $C/T$ peak at 5 K with the application of a magnetic field of 9 T can be interpreted as a manifestation of the field-induced enhancement of Zeeman splitting of the Kramers doublet from less than 2 K to 5 K (see Fig. 3(b)). On the other hand, in the rhombohedral structure phase ($x < 0.6$), the electronic specific heat coefficient $\gamma$ estimated from the fits of $C/T$ (= $\gamma+\beta T^2+\alpha T^4$) at high temperatures increases with increasing $x$ (the detail on analysis is shown in Fig. S6 and Table S6, see Supplementary Material [20]). In addition, the Schottky anomaly due to Kramers doublet is almost absent, reflecting the fact that the valence of the Pr ion is close to +3. Consequently, the increase in $\gamma$ as a function of $x$ below 0.6 indicates an enhancement of the effective mass, as observed for La$_{1-x}$Nd$_x$CuO$_3$ ($x < 0.6$) [16]. As for the origin of the discontinuous jump around $x = 0.6$, it is reasonable to consider the contribution from the Schottky anomaly.

In order to discuss the difference in the band structure between LaCuO$_3$ and PrCuO$_3$, we performed first-principles band calculations as shown in Figs. 4(a) and 4(b), respectively (see Supplementary Material, which includes Refs. [23–31], at [20] for details of first-principles calculations). LaCuO$_3$ has three-dimensional dispersive bands near the Fermi energy. On the other hand, the bandwidth near the Fermi energy of PrCuO$_3$ is narrower than that of LaCuO$_3$, reflecting the low dimensionality and the lattice distortion. Furthermore, several bands, for instance along the Γ-X direction, exhibit almost no dispersion, forming quasi-one-dimensional Fermi surfaces, which reflects the presence of the conductive Cu-O chains along the $c$ axis (see the inset of Fig. 4(b); for the Fermi surfaces at selected energies in Fig. S7, see Supplementary Material [20]). Figures 4(c) and 4(d) depict the density of states (DOS) for LaCuO$_3$ and PrCuO$_3$, respectively. In the case of LaCuO$_3$, the DOS near the Fermi energy is small and consists mainly of Cu-$d$ orbitals. On the other hand, PrCuO$_3$ shows a peak structure in the DOS composed of Pr-$f$ orbitals in the vicinity of the Fermi energy, suggesting the hybridization between the Cu-*d* orbitals and the Pr-*f* orbitals, which is reminiscent of the Kondo systems [32]. Consistent with the calculation, the relatively large Pauli paramagnetism $\chi_0 = 1.2\times10^{-3}$ emu/mol is observed for PrCuO$_3$, which is characteristic to strongly correlated systems [33].

The band calculations largely align with the experimental results. First, the experimental results show that the Cu valence in LaCuO$_3$ is +3, while in PrCuO$_3$, it is closer to +2. The band calculations suggest that the unique divalent nature of Cu ions in PrCuO$_3$ is caused by significant hybridization of the Pr-4$f$ and Cu-3$d$ orbitals. The hybridization of itinerant $d$-orbital and localized $f$-orbital is reminiscent of the "Kondo hybridization" observed in so-called heavy fermion systems. In fact, the band structure presented in Fig. 4(b) shows pseudo-gap structures formed by the hybridization of the Cu-$d$ and Pr-$f$ orbitals (see Fig. S8 in Supplementary Material [20]). The contribution of Pr-$f$ orbitals near the Fermi energy manifests itself as the enhancement of γ by Pr-doping in the rhombohedral phase. The similar orbital hybridization between *A* and *B* sites in perovskite-type oxides has been reported in CaCu$_3$Ru$_4$O$_{12}$, where Kondo hybridization occurs between the localized *d*-orbitals of Cu$^{2+}$ and the delocalized *d*-orbitals of Ru$^{4+}$. However, the electronic properties of CaCu$_3$Ru$_4$O$_{12}$ may be rather different from that of PrCuO$_3$ as it shows metallic behavior [34,35]. Note that a one-dimensional material CeCo$_2$Ga$_8$ also exhibits heavy fermion behavior [36], while coherent metallic conduction is realized even at low temperatures, unlike the case of PrCuO$_3$.

As another origin of the pseudo-gap structure near the Fermi energy, we also consider the possible formation of the Mott-like gap due to the enhancement of electron correlation in the quasi-one-dimensional lattice with nearly divalent Cu ions. However, the inter-site charge transfer is incomplete so that the Pr and Cu ions are in mixed valent states, implying that the energy gap is not fully opened as reflected in the temperature dependence of resistivity and the optical conductivity. In any cases, the emergence of the quasi-one-dimensional structure and the incoherent bands in La$_{1-x}$Pr$_x$CuO$_3$ at high *x* can be associated with the strong *d-f* hybridization and the inter-site charge transfer yielding Jahn-Teller active Cu$^{2+}$ ions.

In conclusion, we found that Pr doping on a metallic perovskite LaCuO$_3$ induces nearly localized nonmetallic state upon the structural transition to a quasi-one-dimensional perovskite phase. This transition is accompanied by significant changes in the valence states of Cu and Pr ions, as well as the electron mass enhancement due to the hybridization between Cu 3$d$, O 2$p$, and Pr 4$f$ orbitals. The structural and electronic transformations suggest the formation of incoherent electronic bands in the Pr-rich compounds, which are further supported by the measurements of optical conductivity and specific heat and the first-principles calculations. These findings highlight the interplay between dimensionality, charge transfer, and *d-f* orbital hybridization in determining the physical properties of La$_{1−x}$Pr$_x$CuO$_3$. This study offers important clues to explore novel correlated states such as a heavy-fermion state and an exotic superconducting state in low-dimensional oxides having multiple cations with charge and orbital degrees of freedom.

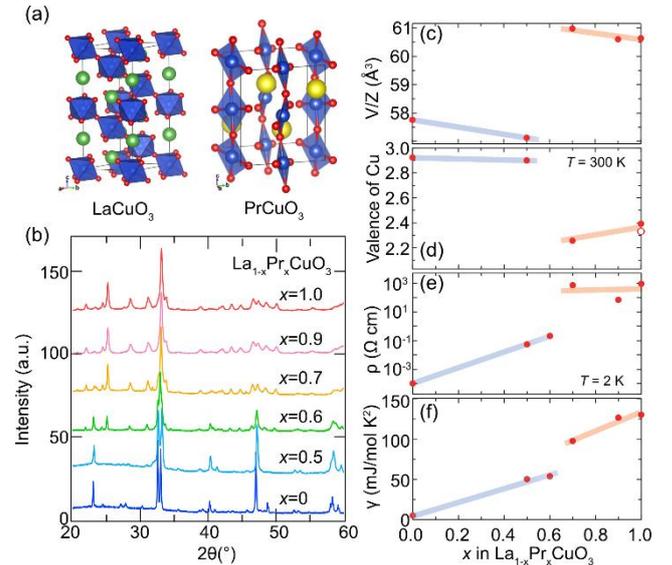

Fig. 1. (a) Crystal structure of LaCuO$_3$ and PrCuO$_3$ [37]. (b) Powder x-ray diffraction patterns of La$_{1-x}$Pr$_x$CuO$_3$ taken at room temperature. (c) Normalized unit cell volume *V/Z*, where *Z* denotes the number of formula units per unit cell, (d) valence of the Cu ion estimated by bond valence sum (filled circles) and x-ray absorption spectra (open circles)[19], (e) resistivity at 2 K, (f) electronic specific heat coefficient γ plotted as a function of *x*. The blue and red lines correspond to the phases with rhombohedral (*R-3c*) and orthorhombic (*Pbnm*) symmetry, respectively.

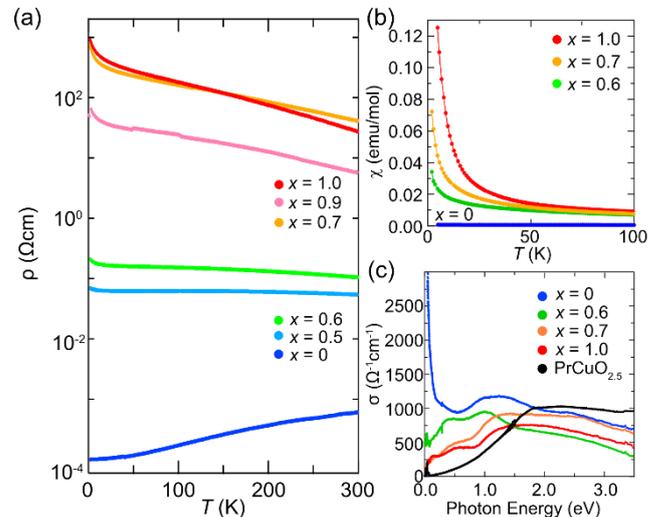

Fig. 2. (a) Temperature dependence of resistivity and (b) magnetic susceptibility for La$_{1-x}$Pr$_x$CuO$_3$ with selected compositions. (c) Optical conductivity spectra at room temperature. The spectra of an oxygen-deficient perovskite PrCuO$_{2.5}$ with insulating behavior is shown as a reference.

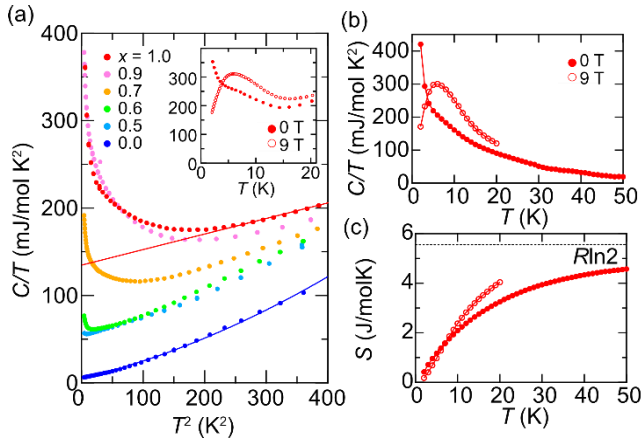

Fig. 3. (a) Specific heat divided by temperature $C/T$ as a function of $T^2$. The blue and red solid lines correspond to the fits of $C/T = \gamma+\beta T^2+\alpha T^4$ and $\gamma+\beta T^2$, respectively; the first term is the electronic contribution, and the second and the third are from the lattice [21]. The inset shows $T$-dependence of $C/T$ of PrCuO$_3$ ($x = 1$) under magnetic fields of 0 T and 9 T. (b) Temperature dependence of background-subtracted $C/T$ and (c) entropy for PrCuO$_3$ measured at 0 T and 9 T. The data of LaCuO$_3$ were adopted as each background.

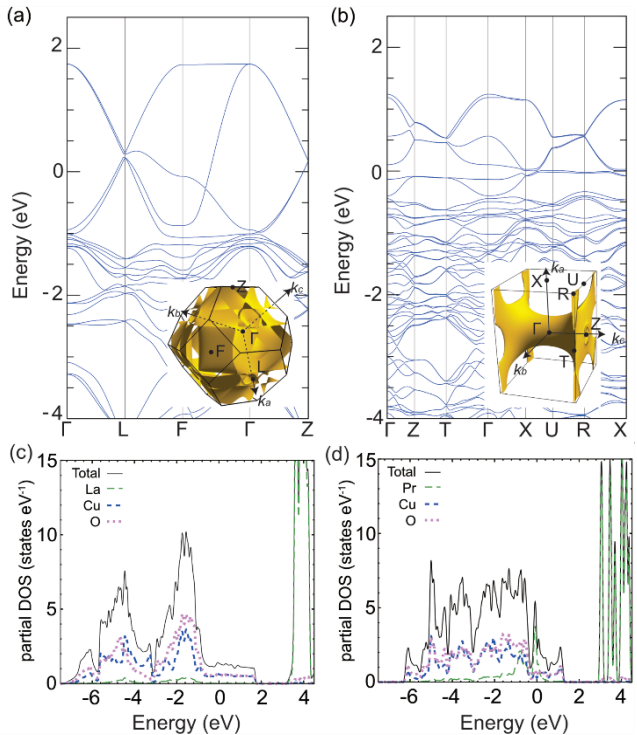

Fig. 4. The band structures for (a) LaCuO$_3$ and (b) PrCuO$_3$. Insets show the Fermi surfaces at 0.5 eV. (c,d) The total density of states (black line) and the partial density of states of La/Pr orbital (green line), Cu orbital (blue line) and O orbital (red line) for (c) LaCuO$_3$ and (d) PrCuO$_3$.

Table 1. Parameters obtained by the Curie-Weiss fitting for La$_{1-x}$Pr$_x$CuO$_3$.

|  | $x = 0.6$ | $x = 0.7$ | $x = 1.0$ |
|---|---|---|---|
| $P_{\text{eff}}$ ($\mu_B$) | 3.44 | 3.42 | 2.68 |
| $\theta$ (K) | -54.3 | -55.0 | -34.9 |
| $\chi_0$ (emu/mol) | 7.8×10$^{-4}$ | 4.2×10$^{-4}$ | 1.2×10$^{-3}$ |

**Acknowledgments**

The authors thank T. Onimaru and K. Matsuhira for useful comments. This work was partly supported by JSPS, KAKENHI (Grants No. 19H05824, 19K14652, 20H01866, 21K18813, 22H01177, 22H00343, 22J13408, 23H04871 and 24K00570), FOREST (JPMJFR236K) and CREST (Grant No. JPMJCR2435) from JST, the Murata Science Foundation and Asahi Glass Foundation. The synchrotron powder XRD was performed with the approvals of the Photon Factory Program Advisory Committee (Proposal No. 2018S2-006 and 2021S2-004).